\documentstyle[sprocl]{article}

\input{psfig}
\def\Journal#1#2#3#4{{#1} {\bf #2}, #3 (#4)}

\def\NPB{{\em Nucl. Phys.} B}
\def\PLB{{\em Phys. Lett.}  B}
\def\PRL{\em Phys. Rev. Lett.}
\def\PRD{{\em Phys. Rev.} D}

\def\be{\begin{equation}}
\def\ee{\end{equation}}
\def\bea{\begin{eqnarray}}
\def\eea{\end{eqnarray}}

\begin{document}

\title{LOOKING FOR EXTRA-DIMENSIONS AT THE WEAK SCALE: EXPERIMENTAL SEARCH
FOR KALUZA-KLEIN STATES SIGNATURES AT THE $e^+e^-$ LINEAR COLLIDER
\footnote{Talk given in the working group
session P6 at the International Workshop on Linear Colliders, Sitges,
Barcelona, Spain, April 28 - May 5, 1999.}  }

\author{ M. BESAN\c CON}

\address{DAPNIA-CEA Saclay, bat. 141, 91191 Gif Sur Yvette, France}


\maketitle\abstracts{An experimental search 
for signatures 
of Kaluza-Klein graviton states at the 500~GeV $e^+e^-$
linear collider, in which the 
graviton states are produced in association with
a photon, is presented. The study of the signal extraction 
is performed with the help of Monte Carlo simulations.}
  
\section{Introduction and motivations}
Quantum gravity is presently best described within the framework of
superstring theories. Superstring~\cite{wit1}, allowing to unify gravity with
the interactions described in the standard model and to remove divergences
from quantum gravity, are known to live in 10 space-time dimensions. 
Furthermore, superstring dualities~\cite{wit2} have tought us that the superstring 
scale 
may not be tied to the Planck scale but becomes a rather arbitrary parameter~\cite{wit3}
which as been proposed to be possibly as small as the TeV scale~\cite{lyk}. 
This observation already opens formally the possibility of the existence
of more than 4 space-time dimensions at the weak scale.
\par
The proposal that the standard model particles and interactions leave in the
usual 4 dimensional space-time while gravity propagates in a higher-dimensional
space~\cite{add} leads to a solution of the hierachy problem. 
In this framework, quantum gravity is characterized by a fundamental scale $M$
of order $\sim$ TeV 
and gravity propagates in a space with $\delta$ extra-dimensions of size R. The
Newtonian gravitationnal constant $G_N$ is then expressed as 
$G_N^{-1} = M_P^2 = 8 \pi R^{\delta} M^{2+\delta}$ where 
$ M_P $ is the Planck mass so that $M$ can be seen as the effective Planck mass
of the higher dimensional theory. 
Such a picture of a standard model confined to a lower dimensional space and gravity
propagating in the bulk is naturally imbedded within 
superstring theories~\cite{aadd}. 
Furthermore, grand unification through extra-dimensions has
been shown to be possible at scales as low as scales close to the weak 
scale~\cite{dudas}.
\par  
These observations lead to a wide spectrum of phenomenological consequences
for conventional Newtonian gravitation, particles physics, astrophysics and 
cosmology~\cite{hewett}. 
In the higher i.e. 3+ $\delta$, dimensional space, the graviton propagates as a 
massless,
spin-2 particle. Projected onto the normal 3 dimensional space, where the standard
model leaves, it appears as a tower of massive Kaluza-Klein excitations.  
\par
In this study, we focus on the production of a Kaluza-Klein (KK) graviton
in association with a photon at $e^+e^-$ colliders, as suggested~\cite{add} for
a possible experimental test, and more specifically at the linear collider
at $\sqrt s = 500 $ GeV. Note that in the early '90, a proposal for a search
for new dimensions at a TeV has been made~\cite{anto}. 
\section{Cross-Sections, Signature and backgrounds}
The cross-section for the process $e^+e^- \rightarrow \gamma $ graviton has been
calculated~\cite{giu} (see also~\cite{hewett}),
without the inclusion of initial state radiation (ISR) of photons,
and yields to:
\begin{equation}\label{eqn:dsig}
 {{d \sigma} \over {dx_{\gamma} d \cos \theta }} (s) =
{\alpha \over 64 } \, 
{2 \pi^{\delta \over 2} \over \Gamma ({\delta \over 2})} \,
({{\sqrt s } \over M})^{\delta+2}  \,
 {1 \over s } f(x_{\gamma},\cos \theta)  
\end{equation}
with
$x_{\gamma}= { 2 E_{\gamma} \over {\sqrt s} }$. The angle
$\theta$ is
the angle between the photon and the beam direction.
In equation~\ref{eqn:dsig}, $ f(x,y)$ is defined by:
\begin{equation}\label{eqn:fxy}
f(x,y)=
{{2(1-x)^{{\delta \over 2} - 1}} \over { x(1-y^2) }} \times
 [ (2-x)^2(1-x+x^2)-3y^2x^2(1-x)-y^4x^4 ]
\end{equation}
The cross-section~\ref{eqn:dsig} has divergences for
for $ x_{\gamma} \rightarrow 0 $ and ${\cos}^2 \theta \rightarrow 1$ which
means that the photon will be close to the beam with an energy spectrum 
favouring very small energies with respect to the beam energy.
\par
\begin{center}
\begin{figure}
\begin{center}
\begin{tabular}{cc}
\psfig{figure=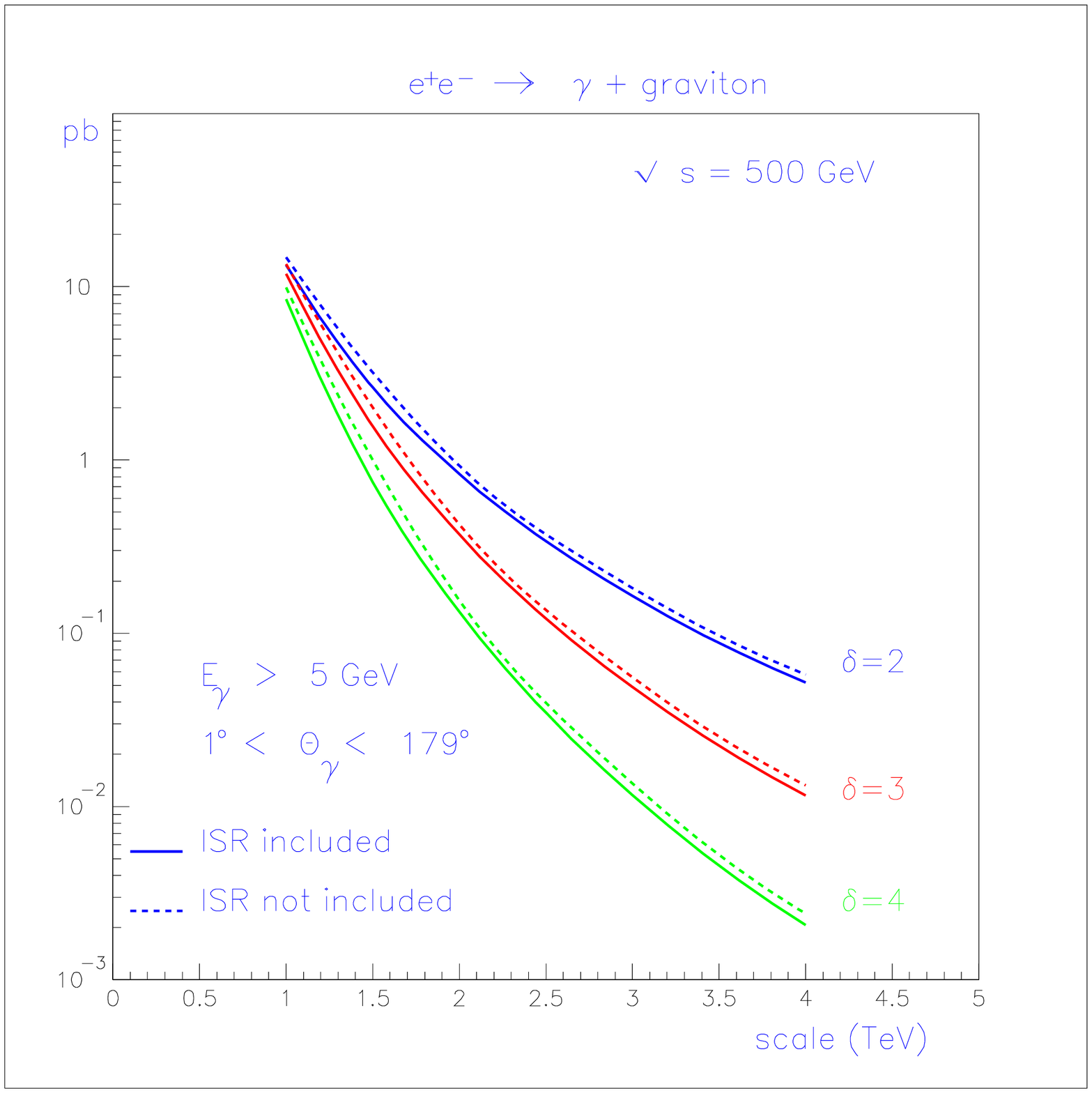,height=2.0in}
\psfig{figure=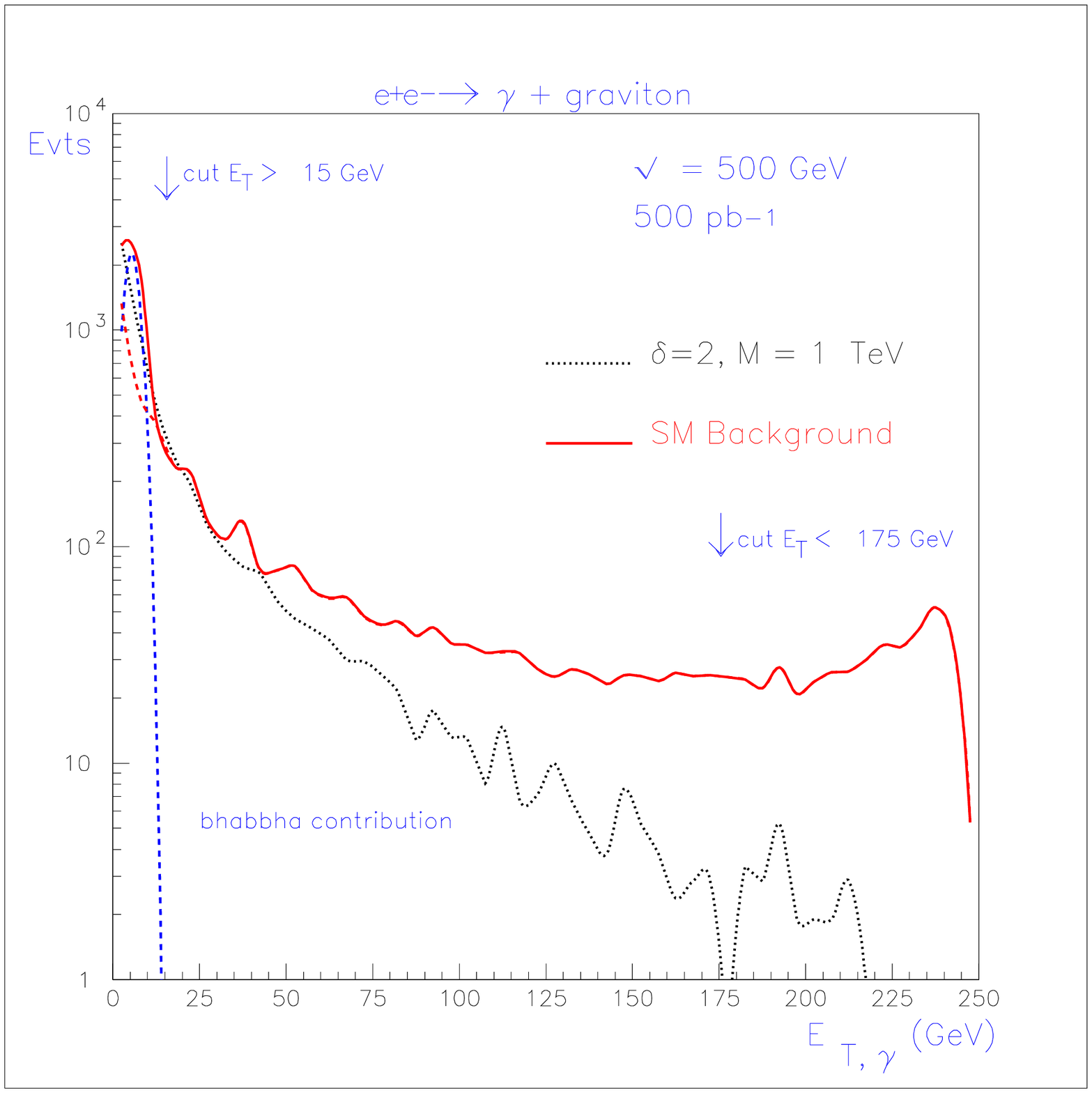,height=2.0in} \\
\end{tabular}
\caption{Left: total cross-section in picobarns.
Right: transverse energy distribution of the candidate photon.
\label{fig:fg1}} 
\end{center}
\end{figure}
\end{center}

In this work, the effect of ISR is included by introducing 
an energy-dependent $e^+e^-$ luminosity function which can read~\cite{pes}:
\begin{equation}\label{eqn:lee}
L_{ee}(z)=[\beta(1-z)^{\beta-1}(1+{3 \over 4} \beta)
- {1 \over 2} (1+z)] \times [1+\alpha_{em}
(  { \pi \over 3} - {1 \over {2 \pi }}) ]
\end{equation}
where:
\begin{equation}\label{eqn:bet}
\beta= {{2 \alpha_{em}} \over \pi } ( \ln {s \over { m_e}} -1)
\end{equation}
In terms of this function, the total cross-section is then given by:
\begin{equation}\label{eqn:rlee}
 {\sigma}(s) = 
\int_0^1 dz L_{ee}(z) {\sigma}(zs) 
\end{equation}
Figure~\ref{fig:fg1} (left part) shows the total cross-sections in the domain
$E_{\gamma} > 5$ GeV and $ 1^o < \theta < 179^o $, with and without
the inclusion of the ISR, as a function of the scale in TeV
for various value of $\delta$ and this at $\sqrt s = $ 500 GeV.
Including the ISR leads to a lowering of the total cross-sections by an amount which
can be of the order of 10~\% or even more, depending on the domain in
$E_{\gamma}$ and $\theta$ considered.
These cross-sections ranges from $10^{-2}$ pb up to several picobarns.  
\par
As the KK graviton G interacts very weakly with matter and has a very long
lifetime, G can be considered as a  
non-interacting stable particle. In consequence, the signature for the process
$e^+e^- \rightarrow \gamma G$ is characterized by the presence of a photon and missing
energy (and, eventually, a photon from ISR). 
\section{Signal extraction}
The main physical backgrounds from processes of the standard model 
for the above signatures come from $\nu \bar {\nu} \gamma (\gamma_{ISR}) $ production
as well as, more marginally, from $Z \gamma$ and $ZZ ({\gamma_{ISR}})$ productions
where the Z boson decays into neutrinos.
Including effects from the detector, such as unefficient measurements or loss 
of particles, 
other backgrounds may become relevant. This may be the case for
Bhabha processes with an ISR photon or   
$\gamma \gamma ({\gamma_{ISR}})$ productions as well as $WW ({\gamma_{ISR}})$,
$We \nu ({\gamma_{ISR}})$, $Zee ({\gamma_{ISR}})$ productions.
In order to study the extraction of a KK graviton signal from these backgrounds
at a typical detector at the linear collider,
we perform a Monte-Carlo study in which the $\nu \bar {\nu} \gamma (\gamma_{ISR}) $ 
events are generated by the NUNUGPV package~\cite{nunu} and the Bhabha
events are generated by the BHWIDE package~\cite{bhwi}.
At $\sqrt s = 500 GeV$, the total
cross-section
for $\nu \bar {\nu} \gamma (\gamma_{ISR}) $ is found to be equal to
9.72 pb with an uncertainty of the order of 20 \% and the total cross-section
for Bhabhas is found to be 14.7 nb. The events corresponding to all the other 
processes quoted above are generated with the help of the PYTHIA 5.7 
package~\cite{pyth}
with the following cross-section at $\sqrt s = 500$~GeV ,
8.2 pb ($Z \gamma$),   
0.55 pb ($ZZ$), 
8.0 pb ($\gamma \gamma $),
7.7 pb ($WW $),
5.3 pb ($We \nu $) and
7.4 pb ($Zee$), all with ISR $\gamma$'s.
\par
We have developped an event generator for the production of a KK graviton
in association with a photon which includes the effect of ISR according
to the above formulae.
\par
All the generated events are then passed through a fast simulation package
of a typical detector at the linear collider i.e. the SIMDET 
package~\cite{simd} in its version 3.1.
\par
The most important parameters for this detectors concern
the electromagnetic calorimeter and the instrumented mask. They have
been tuned (but not yet optimised) such that, for the electromagnetic calorimeter,
the minimum deposited energy is 0.1 GeV, the electron misinterpretation probability
is 0.01, the angular acceptance is 4.5$^o$-175.5$^o$ and the energy resolution
is 10 \%. As for the instrumented mask, we have an angular coverage from
the electromagnetic calorimeter i.e. 4.5$^o$ down to 1$^o$, with a minimum
deposited energy of 10 GeV.
\par 
The events are selected by taking the information from the so called BEST record
of SIMDET 3.1 which gives the best estimate for the energy and direction
of an object. A candidate photon is defined as a detected object having
zero charge and zero mass.
The selection then proceeds by requiring the presence of only one
candidate photon in the event.
Figure~\ref{fig:fg1} (right part) shows the distribution of the transverse energy of
this candidate
photon for all the above backgrounds, assuming an integrated
luminosity of 500 $pb^{-1}$ at $\sqrt s = 500$~GeV. A signal
for 2 extra-dimensions at a scale of 1 TeV is also shown.
The particular contribution of the Bhabha background is singled out
and shown to concern the part below 15 GeV of the 
transverse energy distribution of the candidate photon.
\par
Requiring the transverse energy of the candidate photon to be greater than 15 GeV,
in order to suppress the Bhabha background, and lower than 175 GeV, in order to
reduce the main background from $\nu {\bar {\nu}} \gamma (\gamma_{ISR}) $, 
allows to extract
the signal from KK graviton which will then appear as an excess of events having
a single photon 
and missing energy events.
\par
\begin{center}
\begin{figure}
\begin{center}
\begin{tabular}{c}
\psfig{figure=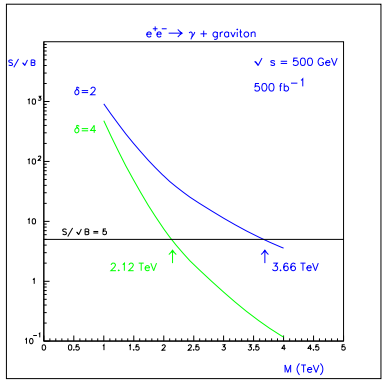,height=2.0in} \\
\end{tabular}
\caption{The $S/ \sqrt B$ ratio.
\label{fig:fg3}}
\end{center}
\end{figure}
\end{center}

Figure~\ref{fig:fg3} shows the $S/ \sqrt B$ ratio, where S stands for signals and 
B for total background, as a function of the scale $M$ extrapolated to
an integrated luminosity of $500 fb^{-1}$. Requiring the $S/ \sqrt B$ ratio to be
greater than 5 allows a reach in terms of M of the order of 3.66 TeV for
2 extra-dimensions and 2.12 TeV for 4 extra-dimensions.
One can try to lower the 15 GeV cut on the transverse energy of the candidate photon
down to 10 GeV or even 5 GeV. This allows to increase the reach in terms of M
up to values of the order of 4 TeV for 2 extra-dimensions but 
the price is a precise control
of the background coming from the low energy part
of the photon spectrum. 
\section{Conclusions and perspectives}
At the $e^+e^-$ linear collider, the search for $e^+e^- \rightarrow \gamma $
graviton seems a promising way to look for extra-dimensions at the TeV scale.
Although the inclusion of the effect of ISR lowers the total cross-sections,
the present Monte-Carlo study, including a fast detector simulation, shows that
this signal can be extracted from the physical and instrumental backgrounds 
and the exploration of the $ M =  3.5$ TeV - 4 TeV mass scale domain for
2 extra-dimensions at $\sqrt s = 500$ GeV at 500 pb$^{-1}$ is feasible.
A very good photon measurement and identification as close to the beams as possible
and a good hermiticity are among crucial requirements
on the detector
at the $e^+e^-$ linear collider for such a measurement.
The study of the beam polarization will be done in a future work and we foresee
an increase of reach in terms M as anticipated in~\cite{hewett} and~\cite{giu}.
Complementary processes such as $e^+e^- \rightarrow Z $
graviton with a Z boson decaying into two fermions
may help in detecting/confirming
the effect of extra-dimensions at the weak scale at the linear collider.
Last but not least, processes such as $e^+e^- \rightarrow \gamma$
dilaton (or even $e^+e^- \rightarrow (\gamma_{ISR})$ dilaton dilaton processes), 
leading also to signature with a single photon and missing energy,
may help in revealing the stringy nature of quantum gravity. Fascinatingly
enough, Kaluza-Klein dilaton production can be distinguished from  Kaluza-Klein
graviton
production since the angular spectrum of the single photon which can
be detected may show competely different features~\cite{ovar}. Work in this direction
is also underway.  
\section*{Acknowledgments}
It a pleasure to thank P. Checchia, M. Spira, S. Tkaczyk, F. Richard,
R. Rueckel and G. Wilson
for discussions
and suggestions on a preliminary
version of this work during the ECFA-DESY Linear Collider Workshop at Oxford.
This work has also benefited from discussions with G. Giudice and J. Wells.
Finally, I. Antoniadis, P. Binetruy, E. Dudas, G. Ovarlez and A. Sagnotti 
deserve special thanks 
for their patience in illuminating and fascinating
explanations on the basics of superstrings (and branes
world) physics.

\end{document}